\documentclass[a4paper,preprintnumbers,showpacs,twocolumn,superscriptaddress,nofootinbib,amsmath,amssymb]{revtex4-1}
\usepackage[dvips]{graphics}
\usepackage[hypertex]{hyperref}
\usepackage{color,hyperref}
\usepackage{times}
\usepackage{mathrsfs}
\hypersetup{
  colorlinks=true,
  linkcolor=blue,
  citecolor=magenta,
  filecolor=magenta,
  urlcolor=blue
}

\def\beq{\begin{equation}}
\def\eeq{\end{equation}}
\def\bear{\begin{eqnarray}}
\def\ear{\end{eqnarray}}
\def\nn{\nonumber}

\usepackage{enumitem}
\usepackage{graphicx,subfigure}% Include figure files
\usepackage{dcolumn}% Align table columns on decimal point
\usepackage{bm}% bold math
\usepackage{color}

\begin{document}

\title{Motion of spinning particles in non asymptotically flat spacetimes}

\author{Bobir Toshmatov}
\email{bobir.toshmatov@nu.edu.kz} \affiliation{Department of
Physics, Nazarbayev  University, 53 Kabanbay Batyr, 010000
Nur-Sultan, Kazakhstan} \affiliation{Ulugh Beg Astronomical
Institute, Astronomicheskaya 33, Tashkent 100052, Uzbekistan}
\affiliation{Tashkent Institute of Irrigation and Agricultural
Mechanization Engineers, Kori Niyoziy 39, Tashkent 100000,
Uzbekistan}

\author{Ozodbek Rahimov}
\email{rahimov@astrin.uz} \affiliation{Ulugh Beg Astronomical
Institute, Astronomicheskaya 33, Tashkent 100052, Uzbekistan}

\author{Bobomurat Ahmedov}
\email{ahmedov@astrin.uz}
\affiliation{Ulugh Beg Astronomical
Institute, Astronomicheskaya 33, Tashkent 100052, Uzbekistan}
\affiliation{Tashkent Institute of Irrigation and Agricultural
Mechanization Engineers, Kori Niyoziy 39, Tashkent 100000,
Uzbekistan}

\author{Daniele Malafarina}
\email{daniele.malafarina@nu.edu.kz} \affiliation{Department of
Physics,  Nazarbayev University, 53 Kabanbay Batyr, 010000
Nur-Sultan, Kazakhstan}

\begin{abstract}

The assumption of asymptotic flatness for isolated astrophysical
bodies may be considered an approximation when one considers a
cosmological context where a cosmological constant or vacuum
energy is present. In this framework we study the motion of
spinning particles in static, spherically symmetric and
asymptotically non-flat spacetimes with repulsive cosmological
vacuum energy and quintessential field. Due to the combined
effects of gravitational attraction and cosmological repulsion,
the region where stable circular orbits are allowed is restricted
by an innermost and an outermost stable circular orbits. We show
that taking into account the spin of test particles may enlarge or
shrink the region of allowed stable circular orbits depending on
whether the spin is co-rotating or counter-rotating with the
angular momentum of the particles.

\end{abstract}

\maketitle

\section{Introduction}\label{sec-intr}

Our current confidence in the existence of astrophysical black
holes relies on the ability of mathematical model to describe
observations. This is true for the gravitational wave signals
observed from binary black hole mergers \cite{GW150914,GW170817}
as well as for the observational evidence based on electromagnetic
radiation coming from accretion disks surrounding the black
holes~\cite{Bambi:APJ} like for example in the case of the Event
Horizon Telescope (EHT) collaboration's image of the supermassive
black hole at the center of the galaxy M87 \cite{EHT}.

All the theoretical models used to describe the observables
obtained from accretion disks rely on the mathematical description
of the motion of test particles and viscous plasma fluid in the
surroundings of the black hole candidate. However, the motion of
test particles may be influenced by a large number of factors,
including, for example, the geometry
\cite{us1,Abdikamalov:PRD:2019,Toshmatov:PRD:2019,Rezzolla:PRD},
the presence of external fields
\cite{Tursunov:2018erf,Stuchlik:2020rls}, the spin of the
particles
\cite{Hojman:PRD:1977,Abramowicz:MNRAS:1979,Suzuki:PRD:1998,Semerak:MNRAS:2007,Plyatsko:PRD:2013}
and possible deviations from classical general relativity
\cite{Hackmann:PRD:2014,Nucamendi:2019qsn}.

In particular, when describing astrophysical isolated bodies one
is usually led to consider asymptotically flat space-times. This
is the result of the assumption that every other gravitational
influence on the astrophysical object can be neglected. However,
in a cosmological context this assumption may not hold as we may
have to take into account the effects of the cosmological
constant. It is well known that stable circular orbits in non
asymptotically flat spacetimes have both an inner and an outer
boundary and the outer boundary may have astrophysical relevance
\cite{Howes,visser}. Similarly, it is well known that circular
orbits are altered once one includes the effects of the spin of
the test particles. The motion of spinning particles in the
different spacetimes with different physical conditions has been
studied by several
authors~\cite{Stuchlik:APS,Semerak:MNRAS:1999,Stuchlik:CQG,Jefremov:PRD:2015,Zhang:PRD:2018,Toshmatov:2019bda}.

%In this paper we shall consider the effects of the presence of
%spin for test particles moving on circular orbits in non
%asymptotically flat spacetimes.

In spite of the fact that the Einstein field equations contain all
the necessary information to derive the equations governing the
motion of a spinning body, given the non linearity of the field
equations, one must necessarily simplify the problem with
additional assumptions. Typically one considers test particles for
which mass and size are negligible as compared to those of the
central object. Within the test particle approximation one can
neglect the effects of gravitational radiation and model the
particle's motion as following geodesics. However, the study of
particles' motion can be improved by considering the spin of the
test particles. In this context, the motion of spinning particles
in general relativity has been widely studied in the pole-dipole
approximation, where the gravitational field and the higher
multipoles of the particle are neglected and the particle itself
is completely characterized by its mass monopole and spin
dipole~\cite{Steinhoff:PRD:2010}. If the particle is
electromagnetically neutral, it interacts with the spacetime only
gravitationally. However, due to the interaction between the spin
of the particle and the curvature of the spacetime, its trajectory
deviates from a geodesics. In such desription the motion of
spinning particles is governed by the Mathisson-Papapetrou-Dixon
equations \cite{Mathisson:1937,PP:1951,Dixon:1970}.

In the setup described above one is led to a set of equations
which is not closed, as there are more unknowns than equations.
This leads to the necessity of introducing some extra condition.
Typically one can introduce the spin-supplementary condition that
ensures that the spin tensor has only three independent
components. There are several variations of the spin-supplementary
condition such as the ones given by
Tulczyjew~\cite{Tulczyjew:1959}, Pirani~\cite{Pirani:1956}, etc.
(for details, see \cite{LG:PRD:2014}.).

On the other hand, when considering real astrophysical phenomena
at large scales, one should take into account cosmological
effects. Measurements of the expansion rate of the universe show
that the expansion is accelerating due to some unknown repulsive
effect which has been given the name of ``dark energy". Such
acceleration can be modeled via the introduction of a cosmological
constant $\Lambda$ in Einstein's equations. This cosmological
constant can in turn be identified as the product of some other
physical effect. For example, in the inflationary paradigm one can
attempt to identify $\Lambda$ with the residual vacuum
energy~\cite{Linde:2005ht}. Similarly, one can interpret $\Lambda$
as resulting from a slowly evolving homogeneous repulsive fluid
called ``quintessence" \cite{Steinhardt:1999nw}. Recent
cosmological tests point out that about 70~$\%$ of the energy
content of the observable universe is due to dark energy and that
the equation of state of dark energy is very close, if not
identical, to that obtained from a repulsive cosmological
constant. Also, numerical estimates based on the cosmic microwave
background indicate the value of the cosmological constant to be
of the order of ${\rm \Lambda\approx10^{-52}}$ ${\rm
m^{-2}}$~\cite{Spergel:2006hy}. Therefore, when studying
astrophysical phenomena at large scales it seems reasonable to
take into account the repulsive effects arising from the presence
of $\Lambda$.

The inclusion of a repulsive cosmological constant into a
spacetime changes significantly its asymptotic structure, as a
black hole, naked singularity, or any compact body becomes
asymptotically de Sitter i.e., not
flat~\cite{Stuchlik1983,Stuchlik:2003dt,Rezzolla:AA:2003,Schee:JCAP,Slany:GRG,Schee:JCAP:2011}.

On the other hand, one could consider a different model for dark
energy, the so called quintessence, for which the accelerated
expansion of the universe is due to a slowly evolving, spatially
homogeneous matter fluid with negative
pressure~\cite{Caldwell:2009zzb,Steinhardt:1999nw}. According to
the idea of quintessence, dark energy in the universe is dominated
by the potential of a scalar field which is still rolling to its
minimum~\cite{Hellerman:2001yi}. This model can be parameterized
by providing the equation of state for the quintessence fluid in
the form $P=\omega \rho$ with $\omega<0$, where the parameter
$\omega$ is in the range of $-1<\omega<-2/3$. One retrieves the
cosmological constant model from the value $\omega=-1$.

In this paper we combine the above mentioned scenarios and study
the motion of spinning particles in the pole-dipole approximation
in non asymptotically flat spacetimes with a dark energy content
described by either a cosmological constant or a quintessence
fluid. The paper is organized as follows: in
section~\ref{sec-basic-eqs} the equations of motion governing the
motion of spinning particle in generic, spherically symmetric
spacetimes is presented. In sections~\ref{sec-SdS}
and~\ref{sec-quint} we apply the derived equations for spinning
particles to the Schwarzschild-de Sitter spacetime and
Schwarzschild spacetime immersed in quintessence, respectively.
Finally, in section~\ref{sec-conclusion} we summarize the obtained
results and discuss the possible implications for astrophysics.
Throughout the paper, we use natural units setting $G=c=1$.

\section{Basic equations}\label{sec-basic-eqs}

In this section we present the general formalism for a spinning
test particle in the Mathisson-Papapetrou-Dixon approximation (up
to the pole - dipole
order)~\cite{Mathisson:1937,PP:1951,Dixon:1970} considering the
spinning particle is moving in the field of a static, spherically
symmetric compact object described by the following
line element:
\bear\label{line-element}
ds^2=-f(r)dt^2+\frac{dr^2}{f(r)}+r^2d{\Omega_2}^2\ , \ear
where $d{\Omega_2}^2=d\theta^2+\sin^2\theta d\phi^2$ is the line
element on the unit 2-sphere and the metric function $f$ depends
only on the radial coordinate $r$.

The dynamics of spinning test particle is governed by the
Mathisson-Papapetrou-Dixon
equation that can be
written as follows:
\bear &&\frac{Dp^\mu}{d\lambda}=-\frac{1}{2} {R^{\mu}}
_{\nu\delta\sigma}u^\nu S^{\delta\sigma}\ ,\label{MPD-eq1}\\
&&\frac{DS^{\mu\nu}}{d\lambda} =2p^{[\mu}u^{\nu]}= p^\mu u^\nu -
p^\nu u^\mu\ ,\label{MPD-eq2} \ear
where $D/d\lambda$ is the covariant derivative along the
particle's trajectory with the affine parameter $\lambda$ given by
$D/d\lambda\equiv u^\mu\nabla_\mu$ and ${R^{\mu}}
_{\nu\delta\sigma}$ is the Riemann tensor. The dynamical
4-momentum and kinematical 4-velocity of the particle are denoted
by $p^\mu$ and $u^\mu$, respectively and the anti-symmetric spin
tensor is denoted by $S^{\mu\nu}$ (with $S^{\mu\nu}=-S^{\nu\mu}$).
Therefore, the spin tensor can have only up to six independent
components. The first equation of motion~(\ref{MPD-eq1}) shows
that the spinning particle does not follow a geodesic trajectory
due to the spin-curvature interaction term ${R^{\mu}}
_{\nu\delta\sigma}u^\beta S^{\delta\sigma}$.

Equations~(\ref{MPD-eq1}) and (\ref{MPD-eq2}) cannot be solved
unless an extra condition, the so called spin-supplementary
condition is introduced. The spin-supplementary condition fixes
the center of the particle and ensures that the spin tensor has
three independent components only. In the literature one can find
several spin-supplementary conditions such as the one proposed by
Tulczyjew~\cite{Tulczyjew:1959}, or the one introduced by
Pirani~\cite{Pirani:1956} (see~\cite{LG:PRD:2014} for details). In
our case, to restrict the spin tensor to generate rotations only,
we employ the so called Tulczyjew spin-supplementary condition
that is given by
\bear\label{SSC-Tulcz}
S^{\mu\nu}p_{\mu}=0\ .
\ear
The Tulczyjew spin-supplementary condition
(\ref{SSC-Tulcz}) implies that the components of the 4-velocity
$u^\alpha$ are determined from the following
relation~\cite{Kunzle:JMP:1972}:
\bear\label{velocity-momentum}
u^\mu = \frac{k}{m^2} \left(p^\mu +
\frac{2S^{\mu\nu}R_{\nu \delta \sigma \rho
}p^{\delta}S^{\sigma\rho }}{4m^2+R_{abcd}S^{ab} S^{cd}}\right)\ ,
\ear
where $k$ is the kinematical mass (or rest mass) of the particle
and it is given by $u^\mu p_\mu=-k$. Moreover, the above
spin-supplementary condition ensures that both the mass, $m$, and
the spin, $S$, of the particle are conserved and are given by
\bear
&&p^{\mu}p_{\mu}=-m^2\ ,\label{mass-conservation}\\
&&S^{\mu\nu}S_{\mu\nu}=2S^2\ .\label{spin-conservation} \ear
Although the momentum (or mass) of the spinning particle is
conserved, the four-velocity of the spinning particle does not
necessarily satisfy the normalization condition $u_\mu u^\mu=-1$,
due to the fact that the four-vectors $p^\mu$ and $u^\mu$ are not
always parallel.

In order to simplify the calculations, hereafter, we consider the
motion of the spinning particles as confined in the equatorial
plane, $\theta=\pi/2$. In general, fixing the value of $\theta$,
reduces the number of independent components of the spin tensor to
three as $S^{\theta\alpha}=0$. Taking this condition together with
the spin-supplementary condition further reduces the number of
independent components of the spin tensor to one. Let this
component be the $S^{tr}$, then by using the Tulczyjew
spin-supplementary condition~(\ref{SSC-Tulcz}), we find the
following relations:
\bear &&S^{t\phi}=-\frac{p_r}{p_\phi}S^{tr}\ ,
\label{SSC-betat}\\
&&S^{r\phi}=\frac{p_t}{p_\phi}S^{tr}\ . \label{SSC-betar}
\ear
Moreover, one has the usual geometry-dependent conserved
quantities associated with the spacetime symmetries via the
Killing vectors, $\xi^\mu$, that can be expressed in the form
\bear\label{killing-conserv}
C_{\xi}=p^\mu\xi_\mu-
\frac{1}{2}S^{\mu\nu}\nabla_\nu\xi_\mu\ .
\ear
For the line element~(\ref{line-element}), which has both axial
and timelike Killing vectors, there are two conserved quantities
for test particles, i.e. energy $E$ and total angular momentum
$L$, and they are given by
\bear
&&{\rm E}= -p_t - \frac{1}{2}f'S^{tr}\ ,\label{energy}\\
&&{\rm L}=p_\phi+rS^{r\phi}\ ,\label{angular-momentum} \ear
where prime denotes the partial derivative with respect to radial
coordinate. From the above equations we can now derive the
equations of motion for spinning test particles. From the
normalization condition~(\ref{mass-conservation}) one finds square
of the radial momentum of the particle as
\bear\label{momentum-r} (p^r)^2 = p_t^2-
f\left(\frac{p_\phi^2}{r^2}+m^2\right)\ . \ear
Now by using the relations (\ref{SSC-betat}), (\ref{SSC-betar}),
and momentum (\ref{momentum-r}), from the spin
conservation~(\ref{spin-conservation}), one can find $S^{tr}$ as
\bear\label{Str} S^{tr}=\frac{s p_\phi}{r}\ , \ear
where $s=S/m$ is specific spin parameter. It should be noted that
$s$ can have both negative and positive values depending on the
direction of spin with respect to direction of $p_\phi$. Finally,
from the conservation of energy~(\ref{energy}) and angular
momentum~(\ref{angular-momentum}), we find the $t$ and $\phi$
components of the four-momentum as
\bear
&&p_t=\frac{-2{\rm E}r-{\rm L}s f'}{2 r-s^2f'}\ ,\label{energy1}\\
&&p_\phi=\frac{2 r ({\rm E}s+{\rm L})}{2r-s^2 f'}\
.\label{angular-momentum1} \ear
One can easily notice that for a particle without spin, i.e.
$s=0$, the momenta corresponding to the time and orbital angular
coordinates are conserved and they are, $p_t=-{\rm E}$, and
$p_\phi={\rm L}$. By inserting covariant momenta (\ref{energy1})
and (\ref{angular-momentum1}) into the radial contravariant
momentum (\ref{momentum-r}), one finds the expression
\bear\label{momentum-r2} \left(p^r\right)^2=A({\rm E}-V_+)({\rm
E}-V_-)\ , \ear
where the coefficient $A$ is given by
\bear\label{notations1} A=\frac{4(r^2-s^2f)}{(2r-s^2f')^2}\ , \ear
and $V_\pm$ are given by
\begin{widetext}
\bear\label{eff-potential} {\rm
V_{\pm}}=\frac{s\left(2f-rf'\right){\rm L}}{2( r^2-s^2
f)}\pm\frac{\left(2r-s^2
f'\right)}{2\left(r^2-s^2f\right)}\sqrt{f\left[{\rm L}^2+m^2
(r^2-s^2f)\right]}\ . \ear
\end{widetext}
One can see from (\ref{momentum-r2}) that in order to have
$(p^r)^2\geq0$, the energy of the particle must satisfy one of the
following conditions: (i) ${\rm E}< V_-$ or (ii) ${\rm E}>V_+$.
Hereafter, we focus on the case of spinning test particle with
positive energy which coincides with the effective potential to be
${\rm V_{eff}}=V_+$.

Now by using the effective potential one can study the
characteristic circular orbits of the spinning test particle in
the field described by the line element
(\ref{line-element}). The particle moves along a circular orbit
in the central field (\ref{line-element}) when the following two criteria are satisfied
simultaneously:
\begin{enumerate}
    \item The particle has zero radial
velocity, i.e.
\bear\label{cond-circular1} \frac{dr}{d\lambda}=0\ , \quad
\Rightarrow \quad {\rm V_{eff}}={\rm E}\ , \ear
    \item The particle has  zero radial acceleration, i.e.
\bear\label{cond-circular2} \frac{d^2r}{d\lambda^2}=0\ , \quad
\Rightarrow \quad \frac{d{\rm V_{eff}}}{dr}=0\ . \ear
\end{enumerate}
By solving eqs. (\ref{cond-circular1}) and
(\ref{cond-circular2}) with respect to the conserved quantities
energy and angular momentum one finds four pairs of
expressions for $s$ and $L$. These four scenarios appear due to the relative orientation of the  spin with respect to the angular momentum (${\rm s\ versus\ L}$) and they
are the following:
\begin{itemize}
\item[(i)] In the case for ${\rm s}>0$ and ${\rm L}>0$ spin and angular momentum are co-rotating;
\item[(ii)] In the case for ${\rm s}>0$ and ${\rm L}<0$ spin and angular momentum are counter-rotating;
\item[(iii)] In the case for ${\rm s}<0$ and ${\rm L}>0$ spin and angular momentum are counter-rotating;
\item[(iv)] In the case for ${\rm s}<0$ and ${\rm L}<0$ spin and angular momentum are co-rotating.
\end{itemize}
Noting the symmetry of the spacetime and looking at the form of
the effective potential~(\ref{eff-potential}), it is easy to see
that each scenario depends only on the sign of the product ${\rm
sL}$. Therefore one can conclude that the above four scenarios
effectively describe only two possible situations, i.e,
co-rotating and counter-rotating spin and angular momentum. Namely
\begin{itemize}
\item[(i)] Co-rotating: ${\rm sL}>0$;
\item[(ii)] Counter-rotating: ${\rm sL}<0$.
\end{itemize}
The analytic expression for the energy and angular momentum of a
spinning particle moving along circular orbit at a fixed radius
$r$ is rather complicated and therefore we shall not report it
here. One must note that in order for the energy of the particle
to be real, the following condition must be satisfied:
\bear\label{st-radius} 8rf'+S^2\left(r^2 f''^2-3
f'^2-6rf'f''\right)+2S^4f'^2f''\geq0\ .\nn\\
\ear
If the spin of the particle is neglected, the condition
(\ref{st-radius}) reduces to $f'\geq0$ and the solution of this
inequality for asymptotically flat spacetimes is just
$r\leq\infty$. However, in the case of non asymptotic flatness the
situation is quite different, as there are repulsive large scale
cosmic effects by the field coupled to gravitation. In this case,
circular orbits can exist only in a region of the spacetime
restricted by a boundary radius, the so-called static radius,
$r_{st}$, that is determined by the equality in
equation~(\ref{st-radius}). At the static radius the gravitational
attraction is just balanced by the cosmic repulsion of the
spacetime.

Another property of circular orbits that is of extreme importance
in astrophysics is their stability. Stability of the orbit is
guaranteed by positivity of the second derivative of the effective
potential with respect to radial coordinate as
\bear\label{cond-isco} \frac{d^2{\rm V_{eff}}}{dr^2}\leq0\ , \ear
where equality corresponds to the smallest allowed
value for the radius of stable circular orbits, namely the innermost stable
circular orbit (ISCO).

One must note that as we have mentioned before, canonical momentum
$p^\mu$ and kinematical velocity $u^\mu$ are not parallel.
Conservation of mass of the spinning particle, $p_\mu p^\mu=-m^2$,
guarantees that the canonical momentum remains timelike along the
trajectory. However, the kinematical four-velocity of the spinning
particle, $u^\mu$, may change from timelike to spacelike.
Of course, a spacelike 4-velocity, i.e. superluminal, motion is physically
meaningless, and therefore the relation $u_\mu u^\mu>0$ must not be allowed for
real particles, thus imposing an extra condition.
In fact, before $u_\mu u^\mu$ becomes positive the particle must cross the
boundary between timelike and spacelike trajectories where the
relation $u_\mu u^\mu=0$ holds. Therefore, in order to keep
the motion of the spinning particle from becoming spacelike, one must impose that
the condition $u_\mu u^\mu=0$ is not achieved on the whole trajectory. The
kinematical four-velocity and dynamical momentum relation depends
strongly on the spin-supplementary condition. In the Tulczyjew
spin-supplementary condition~(\ref{SSC-Tulcz}) this relation is
given by (\ref{velocity-momentum}). Let us now consider the
superluminal limit for a spinning particle moving on
circular orbit. For simplicity, we introduce the notation vector $v^\mu$
given by
\bear\label{vmu}
v^\mu=\frac{2S^{\mu\nu}R_{\nu \delta \sigma \rho
}p^{\delta}S^{\sigma\rho }}{4m^2+\Delta}\ , \ear
where
\bear\label{notation-delta} \Delta\equiv
R_{abcd}S^{ab}S^{cd}=\frac{2S^2 }{r^2} \left(f''p_\phi^2-\frac{r
f'}{f}p_t^2\right)\ , \ear
so that the kinematical four-velocity is written in the form of
$u^\mu=p^\mu/m+v^\mu$. The components of vector $v^\mu$ for a
spinning particle moving on a circular orbit are given by
\bear v^\mu= \frac{2S^2 p_t p_\phi \left(f'-r
f''\right)}{r^3f\left(\Delta+4 m^2\right)}\left\{-p_\phi,0,0,p_t\right\}\ ,\nonumber\\
\ear
Then in terms of the spacetime with line element~(\ref{line-element})
equation $u_\mu u^\mu=0$ takes the following form:
\bear\label{superluminal1}
\left[\frac{2S^2p_t p_\phi\left(f'-r
f''\right)}{r^2 \left(\Delta+4 m^2\right)}\right]^2=f\ , \ear
The above relation confirms the fact that in the case of absence
of spin of the particle, i.e. $S=0$, the particle crosses the
superluminal bound at the horizon of the spacetime that is
determined by the equation $f=0$. One should note here that the
superluminal limit of the spinning particle is spin-supplementary
condition dependant, i.e., with switching to another spin-supplementary
condition, one would have different equation for the superluminal limit
than (\ref{superluminal1}). For example, if the 4-velocity, $u^\mu$, is
parallel to the 4-momentum, $p^\mu$, ($u^\mu \| p^\mu$) spin-supplementary
condition is applied, the spinning particle would cross the superluminal
limit at the horizon of the black hole as in the case of non-spinning
particle~\cite{Semerak:MNRAS}.

Finally, there is one more important quantity that
characterizes the particle and is worth discussing.
This is the particle's angular velocity on the circular
orbit and it is defined by
\bear\label{omega}
\Omega=\frac{u^\phi}{u^t}\ .
\ear
In the following sections we apply the above setup to two line elements,
one describing the Schwarzschild-de Sitter spacetime and the other
describing a black hole in an expanding universe with quintessence.
Given the complicated nature of the equations most results will be obtained from numerical calculations.

\section{Spinning particle in the Schwarzschild-de Sitter
spacetime}\label{sec-SdS}

The line element of the Schwarzschild-de Sitter spacetime is given
by (\ref{line-element}) with the metric function
\bear\label{sds-metric-function}
 f(r)=1-\frac{2M}{r}-\frac{\Lambda}{3}r^2\ ,
\ear
where $M$ is the total mass of the black hole, while $\Lambda$ is a
cosmological constant. For convenience, let us use dimensionless coordinates
by redefining $t$ and $r$ as
$t/M\rightarrow t$, $r/M\rightarrow r$ and introducing the
dimensionless cosmological parameter in place of the
cosmological constant as:
\bear \lambda=\frac{\Lambda}{3}M^2\ .
\ear
The main properties of the Schwarzschild-de Sitter spacetime have
been widely studied in the literature (-- see
Refs.~\cite{Stuchlik:1999qk,Toshmatov:EPJP}), however,
for the sake of our further calculations, it is useful to present
some crucial results here. For $0<\lambda<1/27$, the
Schwarzschild-de Sitter black hole spacetime has two coordinate
singularities given by $f=0$, which indicate the event and
cosmological horizons. They are located at
\bear\label{hor}
&&r_h=\frac{2}{\sqrt{3\lambda}}\cos\frac{\pi+\arccos(3\sqrt{3\lambda})}{3}\ ,\\
&&r_c=\frac{2}{\sqrt{3\lambda}}\cos\frac{\pi-\arccos(3\sqrt{3\lambda})}{3}\
, \ear
The regions inside event horizon, $r<r_h$, and beyond cosmological
horizon, $r>r_c$, are dynamic. Therefore, we consider the static
region between these two horizons, $r_h<r<r_c$ where particle
motion is timelike.

For $\lambda=1/27$, the event and cosmological horizons merge into
one degenerate horizon which coincides with the unstable lightring
and is located at $r_h=r_c=3$. For $\lambda>1/27$, the
Schwarzschild-de Sitter spacetime represents a naked singularity.
\begin{figure}[h]
\centering
\includegraphics[width=0.48\textwidth]{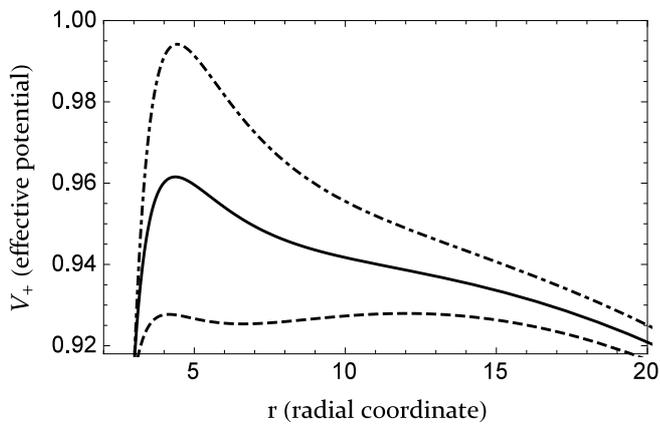}
\caption{\label{fig-eff-SdS} Qualitative picture of the radial profile
of the effective potential for spinning particles with fixed angular
momentum in the Schwarzschild-de Sitter spacetime for a given value of the
cosmological parameter. Dot-dashed, solid, and dashed curves
correspond to spinning particles with positive ($s>0$),
zero ($s=0$), and negative ($s<0$) spin, respectively. }
\end{figure}

\begin{figure}[t]
\centering
\includegraphics[width=0.48\textwidth]{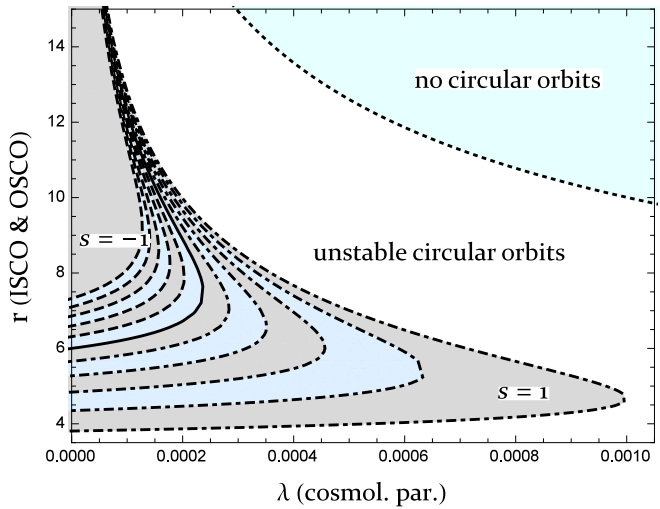}
\caption{\label{fig-isco-SdS} Radii of marginally stable circular
orbits as a function of the cosmological parameter $\lambda$ for
different values of the spin of the test particle. For a given
value of $\lambda$ the intersection with the lower (upper) part of
the curve determines the ISCO (OSCO). Dashed, solid, and dotdashed
curves correspond to the cases of spinning particle with negative
($s<0$), zero ($s=0$), and positive ($s>0$) spins. The dotted
curve represents the static radius. The shaded regions correspond
to regions where stable circular orbits can exist, while in the
white region between OSCO and the static radius circular orbits
exist but are unstable. Beyond the static radius (in cyan region),
circular orbits cannot exist.}
\end{figure}

The motion of spinning particles in the Schwarzschild-de Sitter
spacetime has been studied before
in~\cite{Mortazavimanesh:2009rm,Kunst:SdS}. Here we will present a
more detailed analysis, to gain some general insight on different
non asymptotically flat spacetimes. As we know, the motion of test
particles in a central field is governed by the effective
potential. Therefore, before studying the motion of the spinning
particles, it is important to know the form of the effective
potential and the effects of the spin parameter. To this aim, in
Fig.~\ref{fig-eff-SdS} we present a qualitative picture of the
effective potential for the spinning particle with positive, zero
and negative spin in the field of Schwarzschild-de Sitter
spacetime. It can be seen that an increase in the value of the
spin raises the height of effective potential. Moreover, the
existence of two local maxima in the effective potential indicates
that the existence of bound orbits occurs only between these two
maxima. In asymptotically flat spacetimes circular orbits can
exist at every radius outside the ISCO. However, in asymptotically
non-flat spacetimes such as the Schwarzschild-de Sitter one,
circular orbits can exist only in a region bounded between the
ISCO and another radius, where the cosmological repulsive vacuum
effect is balanced by the gravitational attraction of central
objects. This radius is determined by the equality holding in
(\ref{st-radius}), which, for the Schwarzschild-de Sitter case
takes the form
\bear\label{st-radius-SdS}
4 r^6 \left(-1+\lambda r^3\right)&&+r^3
S^2\left(-13-4 \lambda r^3+8 \lambda^2r^6\right)\nn\\
&&+4 S^4 \left(2-3 \lambda r^3+\lambda^3r^9\right)=0\ .
\ear
The solution of equation (\ref{st-radius-SdS}) is very sensitive
to the cosmological parameter, while a change in the spin of the
particle does not effect much the value of the static radius.
However, this equation cannot be solved analytically. Therefore,
in order for the solution of equation (\ref{st-radius-SdS}) to be
more informative, we included it (dotted curve) into
Fig.~\ref{fig-isco-SdS}.

On the other hand, the stability of circular orbits is one of the
most important characterizations of astrophysical spacetimes. As
we have mentioned in the previous section, the stability is
guaranteed by the condition (\ref{cond-isco}). Unlike the case of
asymptotically flat spacetimes, in the Schwarzschild-de Sitter
metric the region where stable circular orbits can exist is
bounded not only by an inner radius, called innermost stable
circular orbit (ISCO), but also by outer radius that is called
outermost stable circular orbit (OSCO). Non-spinning particles can
have stable circular orbits around Schwarzschild-de Sitter black
holes only if values of the cosmological parameter is in the
following range: $\lambda\in[0,4/16875]$ (that corresponds to
$\Lambda\in[0,4/5625M^2]$) (solid curve in
Fig.~\ref{fig-isco-SdS}). If the value of cosmological parameter
is equal to the upper limit, $\lambda=4/16875$, particles can have
only one stable orbit where the ISCO and OSCO coincide. For larger
values of $y$ there are no stable orbits anymore anywhere in the
spacetime. As it was shown in Fig.~\ref{fig-isco-SdS}, the
inclusion of the particle's spin can expand or shrink the range of
$y$ for which circular orbits exist, depending on the sign of
product of spin and angular momentum of the particle.

All circular orbits in the region between OSCO and the static
radius (in Fig.~\ref{fig-isco-SdS} white region) are unstable.
Increasing the value of the spin causes the ISCO to tend towards
the horizon of the black hole. However, before reaching the event
horizon, it crosses superluminal motion limit. Therefore, for
increasing values of the spin parameter, the stable circular
orbits change from being bounded by the superluminal motion
radius, ISCO, and OSCO to be bounded by superluminal motion radius
and OSCO only. Regarding the effect of the spin on the OSCO, an
increase in the value of the spin pushes the OSCO towards the
static radius. Since the static radius is not very sensitive to
changes in the value of the spin, increase in the value of the
spin causes an increase in the region where stable circular orbits
are allowed.
\begin{table}[h]
\caption{\label{tab-superl-SdS} Values of the characteristic
parameters of spinning test particles with superluminal bound
($u_\mu u^\mu=0$) at the ISCO of the Schwarzschild-de Sitter black
hole. Where ${\rm l_{ISCO}}$=${\rm L_{ISCO}}+{\rm s(threshold)}$.}
\begin{ruledtabular}
\begin{tabular}{cccccc}
$\lambda$ & ${\rm s(threshold)}$ & ${\rm r_{ISCO}(min)}$ &  ${\rm
E_{ISCO}}$ & ${\rm l_{ISCO}}$ & ${\rm \Omega_{ISCO}}$ \\ \hline
 0.035 & 7.4950 & 2.9118 & 0.1396 & 6.2627 & 0.0440 \\
 0.030 & 4.0494 & 2.7827 & 0.2743 & 2.4301 & 0.0795 \\
 0.025 & 3.0703 & 2.7037 & 0.3753 & 1.1601 & 0.1030 \\
 0.020 & 2.5508 & 2.6485 & 0.4651 & 0.3917 & 0.1221 \\
 0.015 & 2.2153 & 2.6074 & 0.5494 & 0.1684 & 0.1388 \\
 0.010 & 1.9761 & 2.5756 & 0.6308 & 0.6159 & 0.1539 \\
 0.005 & 1.7948 & 2.5503 & 0.7106 & 0.9932 & 0.1679 \\
 0.000 & 1.6518 & 2.5299 & 0.7894 & 1.3226 & 0.1809 \\
\end{tabular}
\end{ruledtabular}
\end{table}
\begin{figure}[h]
\centering
\includegraphics[width=0.48\textwidth]{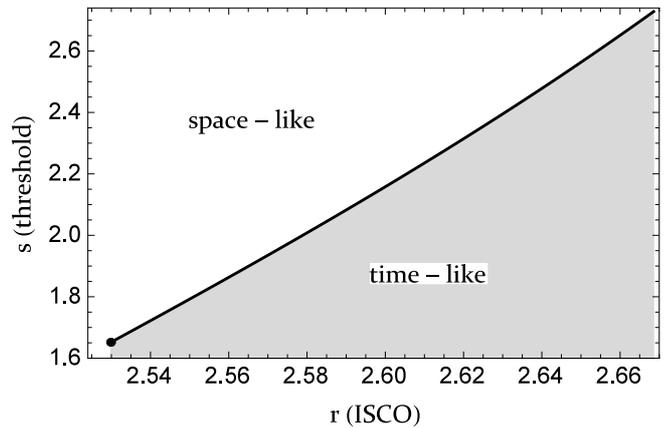}
\caption{\label{fig-superl-SdS} Relation between the threshold
value for the spin of test particles and the radius of the ISCO in
the Schwarzschild-de Sitter spacetime depending on the
cosmological parameter $\lambda\in[0,1/27]$. The solid line
corresponds to the solution of the superluminal limit relation
$u_\mu u^\mu=0$, with the black point at the bottom left
corresponding to the superluminal limit of the spinning particle
at the ISCO of the Schwarzschild black hole ($\lambda=0$).}
\end{figure}

As said, one must ensure that the particle's motion is always
time-like. Therefore it is important to consider the allowed
limits for the parameters of spinning test particles moving along
ISCO and OSCO around Schwarzschild-de Sitter black hole due to the
existence of the superluminal bound. Such parameters are the
particle's energy, angular momentum and angular velocity at the
ISCO, labelled as ${\rm E_{ISCO}}$, ${\rm L_{ISCO}}$ and ${\rm
\Omega_{ISCO}}$. To find these values, one must solve the
equations~(\ref{cond-isco}) and (\ref{st-radius-SdS}),
simultaneously. In Tab.~\ref{tab-superl-SdS} and
Fig.~\ref{fig-superl-SdS} are shown the values of the parameters
for a spinning test particle with superluminal limit ($u_\mu
u^\mu=0$) at the ISCO of the Schwarzschild-de Sitter black hole.
One can see from Tab.~\ref{tab-superl-SdS} that as the influence
of cosmological repulsion increases, the superluminal limits of
radius of ISCO and spin of the particle increase. However, the
corresponding energy and angular velocity of the particle
decrease.

In Tab.~\ref{tab-isco-SdS} we present some values for the
characteristic parameters of a spinning particle at the ISCO and
OSCO of the Schwarzschild-de Sitter spacetime for different values
of spin and cosmological parameter $\lambda$. A particle moving
along the ISCO always has smaller energy and angular momentum than
one with the same spin moving along the OSCO of the
Schwarzschild-de Sitter spacetime. Of course, the values
corresponding to $\lambda=0$ represent the ones of the
Schwarzschild spacetime, i.e. if the cosmological constant is
absent, the radius of the OSCO shifts to infinity, the particle at
the OSCO can be considered as rest and its angular velocity
becomes negligible.
\begin{table*}[t]
\caption{\label{tab-isco-SdS} Values of the characteristic parameters
of spinning particles at ISCO and OSCO of the Schwarzschild-de
Sitter black hole.}
\begin{ruledtabular}
\begin{tabular}{cccccccccccc}
${\rm s}$ & $\lambda$ & ${\rm r_{ISCO}}$ &  ${\rm r_{OSCO}}$ &
${\rm E_{ISCO}}$ & ${\rm E_{OSCO}}$ & $\rm L_{{\rm ISCO}}$ & ${\rm
L_{OSCO}}$ &  ${\rm \Omega_{ISCO}}$
& ${\rm \Omega_{OSCO}}$ & ${\rm u^2_{ISCO}}$ & ${\rm u^2_{{\rm OSCO}}}$\\
\hline
     & 0.0009 & 4.2972 & 5.2746 & 0.8825 & 0.8832 & 2.5604 & 2.5690 & 0.0927 & 0.0681 & -0.9992 & -0.9998 \\
     & 0.0007 & 4.0964 & 6.0470 & 0.8883 & 0.8923 & 2.5874 & 2.6399 & 0.1001 & 0.0558 & -0.9987 & -0.9999 \\
1.0  & 0.0005 & 3.9819 & 7.0077 & 0.8938 & 0.9037 & 2.6113 & 2.7498 & 0.1049 & 0.0451 & -0.9983 & -0.9999 \\
     & 0.0003 & 3.8997 & 8.5721 & 0.8992 & 0.9190 & 2.6332 & 2.9405 & 0.1087 & 0.0337 & -0.9979 & $\approx$-1 \\
     & 0.0000 & 3.8073 & $\infty$ & 0.9069 & 1 & 2.6636 & $\infty$ & 0.1133 & 0 & -0.9973 & $\approx$-1 \\ \hline
    & 0.00035 & 5.7268 & 7.2707 & 0.9171 & 0.9178 & 3.0580 & 3.0700 & 0.0668 & 0.0456 & -0.9999 & $\approx$-1 \\
    & 0.0003  & 5.5384 & 7.9438 & 0.9194 & 0.9215 & 3.0743 & 3.1134 & 0.0706 & 0.0397 & -0.9999 & $\approx$-1 \\
0.5 & 0.0002  & 5.3160 & 9.5612 & 0.9238 & 0.9309 & 3.1033 & 3.2493 & 0.0757 & 0.0299 & -0.9999 & $\approx$-1 \\
    & 0.0001 & 5.1715 & 12.5560 & 0.9279 & 0.9449 & 3.1293 & 3.5356 & 0.0794 & 0.0198 & -0.9999 & $\approx$-1 \\
    & 0.0000 & 5.0633 & $\infty$ & 0.9319 & 1 & 3.1533 & $\infty$ & 0.0824 & 0 & -0.9999 & $\approx$-1 \\ \hline
    & 0.0002 & 6.1322 & 9.2165 & 0.9294 & 0.9320 & 3.2807 & 3.3361 & 0.0630 & 0.0324 & $\approx$-1 & $\approx$-1 \\
0.2 & 0.0001 & 5.8380 & 12.3784 & 0.9345 & 0.9453 & 3.3187 & 3.5973 & 0.0687 & 0.0205 & $\approx$-1 & $\approx$-1 \\
    & 0.0000 & 5.6562 & $\infty$ & 0.9392 & 1 & 3.3521 & $\infty$ & 0.0727 & 0 & $\approx$-1 & $\approx$-1 \\ \hline
    & 0.0002 & 6.7224 & 8.8917 & 0.9319 & 0.9327 & 3.3768 & 3.3960 & 0.0556 & 0.0350 & -1 & -1 \\
0.0 & 0.0001 & 6.2425 & 12.2499 & 0.9376 & 0.9456 & 3.4243 & 3.6388 & 0.0633 & 0.0211 & -1 & -1 \\
    & 0.0000 & 6 & $\infty$ & 0.9428 & 1 & 3.4641 & $\infty$ & 0.0680 & 0 & -1 & -1 \\ \hline
     & 0.0002 & 7.6011 & 8.2476 & 0.9336 & 0.9337 & 3.4583 & 3.4588 & 0.0463 & 0.0403 & $\approx$-1 & $\approx$-1 \\
-0.2 & 0.0001 & 6.6256 & 12.1115 & 0.9400 & 0.9460 & 3.5181 & 3.6804 & 0.0588 & 0.0217 & $\approx$-1 & $\approx$-1 \\
     & 0.0000 & 6.3114 & $\infty$ & 0.9457 & 1 & 3.5645 & $\infty$ & 0.0643 & 0 & $\approx$-1 & $\approx$-1 \\ \hline
     & 0.00015 & 7.6978 & 9.5842 & 0.9394 & 0.9397 & 3.6067 & 3.6161 & 0.0469 & 0.0323 & $\approx$-1 & $\approx$-1 \\
-0.5 & 0.0001 & 7.1763 & 11.8802 & 0.9429 & 0.9465 & 3.6420 & 3.7434 & 0.0532 & 0.0228 & $\approx$-1 & $\approx$-1 \\
     & 0.0000 & 6.7294 & $\infty$ & 0.9492 & 1 & 3.6985 & $\infty$ & 0.0598 & 0 & $\approx$-1 & $\approx$-1 \\
\end{tabular}
\end{ruledtabular}
\end{table*}

\section{Spinning particle in the Schwarzschild spacetime with
quintessence}\label{sec-quint}

We now turn the attention to the case of a black hole immersed in
quintessence and perform a similar analysis to the one presented
in the previous section. The simplest solution of this kind was
suggested by Kiselev~\cite{Kiselev:2002dx} and it is given by the
line element~(\ref{line-element}) with the following metric
function:
\bear\label{metric-function-quint}
f(r)=1-\frac{2M}{r}-cr^{-1-3\omega}\ ,
\ear
where $M$ is the gravitational mass of the black hole and $c$ is
the quintessential parameter representing the strength of the
quintessential field around the black hole. The parameter $\omega$
is related to the dark energy equation of state (EoS). The
equation of state relating the pressure, $P$ to the energy
density, $\rho$, of the quintessential field is assumed to be
linear, in agreement with standard cosmological model. Thus we set
$P=\omega\rho$ and in order for the EoS to describe a dark energy
fluid the values of $\omega$ must be chosen in the following
range: $\omega\in(-1;-1/3)$.

In the following, for simplicity, we will focus on the commonly
used value $\omega=-2/3$. In this case, the metric
function~(\ref{metric-function-quint}) takes the simple form:
\bear\label{metric-function-quint1} f(r)=1-\frac{2M}{r}-cr\ .
\ear
As it was shown in~\cite{Toshmatov:2015npp}, this spacetime
represents either a black hole or a naked singularity depending on
the values of the parameters $c$. If the value of quintessence
parameter is in the range of $0<c\leq 1/8M$, the spacetime
represents the black hole with two horizons, the event horizon and
a cosmological `quintessential' horizon. The horizons are located
at
\bear
&&r_{h}=\frac{1-\sqrt{1-8cM}}{2c}\ ,\\
&&r_q=\frac{1+\sqrt{1-8cM}}{2c}\ .
\ear
In the region between these two horizons, the spacetime is static,
outside this region, it is dynamic. Therefore, we hereafter
consider the particle's motion in the static region. If the value
of the quintessence parameter is $c=1/8M$, similarly to the
Schwarzschild-de Sitter case, both horizons merge into one
degenerate horizon at $r_h=r_q=4M$. If the value of quintessence
parameter is $c>1/8M$ no horizon is present and the spacetime
represents a naked singularity.

As this spacetime is also not asymptotically flat, similarly to
the Schwarzschild-de Sitter case, we find that there exists a
static radius here as well. The static radius is the obtained by
solving the following equation:
\bear\label{st-radius-quint} 8\left(\frac{2M}{r}-cr\right)&&+S^2
\left(-3 c^2-\frac{12 c M}{r^2}+\frac{52
M^2}{r^4}\right)\nonumber\\&&-\frac{8 M
S^4}{r^5}\left(\frac{2M}{r}-cr\right)^2=0\ .
\ear
Again, circular orbits are bounded by the static radius. However,
not all circular orbits are stable. For a non spinning particle to
have stable circular orbits in this spacetime, the value of the
quintessence parameter must be in the following range:
$c\in[0;(3-2\sqrt{2})/32M]$ and stable circular orbits are bounded
by an inner radius at the ISCO and by an outer radius at the OSCO.
The dependence of radii of stable circular orbits on the spin of
the particle and the quintessence parameter is qualitatively
similar to the Schwarzschild-de Sitter spacetimes, therefore we
will not repeat the whole analysis here.

In Tab.~\ref{tab-superl-quint} we list the values of the characteristic parameters
${\rm E_{ISCO}}$, ${\rm L_{ISCO}}$ and ${\rm \Omega_{ISCO}}$ for a
spinning particle with superluminal limit, moving along
stable circular orbits in the Schwarzschild spacetime with
quintessence for several values of the quintessence
parameter.
\begin{table}
\caption{\label{tab-superl-quint} The same as
Tab.~\ref{tab-superl-SdS} for the case of a Schwarzschild black hole
with quintessence.}
\begin{ruledtabular}
\begin{tabular}{cccccc}
$c$ & ${\rm s(threshold)}$ & ${\rm r_{ISCO}(min)}$ &  ${\rm
E_{ISCO}}$ & ${\rm l_{ISCO}}$ & ${\rm \Omega_{ISCO}}$ \\ \hline
 0.035 & 1.9807 & 2.6983 & 0.6756 & 0.9861 & 0.1502 \\
 0.030 & 1.9226 & 2.6712 & 0.6935 & 1.0489 & 0.1549 \\
 0.020 & 1.8193 & 2.6203 & 0.7275 & 1.1573 & 0.1638 \\
 0.015 & 1.7731 & 2.5964 & 0.7437 & 1.2043 & 0.1682 \\
 0.010 & 1.7300 & 2.5734 & 0.7593 & 1.2473 & 0.1725 \\
 0.005 & 1.6897 & 2.5512 & 0.7746 & 1.2866 & 0.1767 \\
 0.002 & 1.6667 & 2.5383 & 0.7835 & 1.3086 & 0.1792 \\
 0.001 & 1.6592 & 2.5341 & 0.7864 & 1.3156 & 0.1801 \\
 0.000 & 1.6518 & 2.5299 & 0.7894 & 1.3226 & 0.1809 \\
\end{tabular}
\end{ruledtabular}
\end{table}
\begin{figure}[h]
\centering
\includegraphics[width=0.48\textwidth]{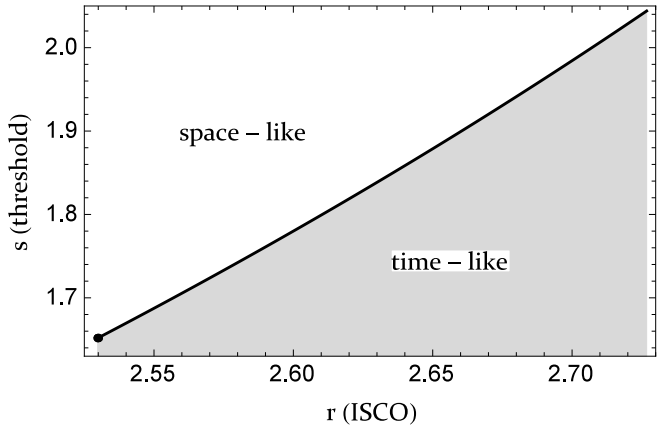}
\caption{\label{fig-superl-quint} The same as
Fig.~\ref{fig-superl-SdS} for the case of a Schwarzschild black hole
with quintessence. The solid line corresponds to the
values of the quintessence parameter in the range $c\in[0,
0.043]$.}
\end{figure}

For small values of the spin parameter, stable circular orbits are
bounded by the ISCO and OSCO, while the superluminal limit radius
is located below the ISCO and just above the event horizon.
Increasing the value of the spin of the particle, the radius of ISCO
approaches the event horizon, while the superluminal limit
radius tends to the ISCO. At some point, for an intermediate
values of the spin parameter
%{\bf (DM: can we put an estimate of the value of $s$ here?)}
they meet and after that, as the value of the spin increases,
the stable circular orbits are restricted by superliminal
limit radius, ISCO and OSCO. Finally, for the large values of the spin
parameter the stable circular orbits are bounded by the
superluminal limit radius and OSCO only.
\begin{figure}[h]
\centering
\includegraphics[width=0.48\textwidth]{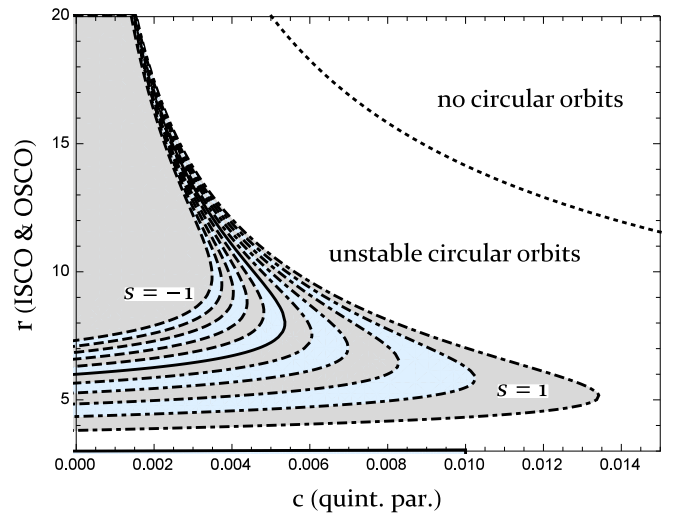}
\caption{\label{fig-isco-quint} The same as
Fig.~\ref{fig-isco-SdS} for the case of a Schwarzschild black hole
with quintessence. }
\end{figure}
\begin{table*}[t] \caption{\label{tab-isco-quint} The same as
Table~\ref{tab-isco-SdS} for the case of a Schwarzschild black hole
with quintessence.}
\begin{ruledtabular}
\begin{tabular}{cccccccccccc}
${\rm s}$ & $c$ & ${\rm r_{ISCO}}$ &  ${\rm r_{OSCO}}$ & ${\rm
E_{ISCO}}$ & ${\rm E_{OSCO}}$ & $\rm L_{{\rm ISCO}}$ & ${\rm
L_{OSCO}}$ &  ${\rm \Omega_{ISCO}}$
& ${\rm \Omega_{OSCO}}$ & ${\rm u^2_{ISCO}}$ & ${\rm u^2_{{\rm OSCO}}}$\\
\hline
     & 0.013 & 4.7968 & 5.5974 & 0.8533 & 0.8535 & 2.5860 & 2.5887 & 0.0768 & 0.0604 & -0.9997 & -0.9999 \\
     & 0.010 & 4.3223 & 7.0328 & 0.8670 & 0.8718 & 2.6167 & 2.6900 & 0.0912 & 0.0429 & -0.9992 & $\approx$-1 \\
1.0  & 0.006 & 4.0537 & 9.6210 & 0.8838 & 0.9016 & 2.6423 & 2.9534 & 0.1017 & 0.0270 & -0.9985 & $\approx$-1 \\
     & 0.003 & 3.9172 & 14.0813 & 0.8956 & 0.9312 & 2.6549 & 3.4081 & 0.1079 & 0.0154 & -0.9979 & $\approx$-1 \\
     & 0.000 & 3.8073 & $\infty$ & 0.9069 & 1 & 2.6636 & $\infty$ & 0.1133 & 0 & -0.9973 & $\approx$-1 \\ \hline
     & 0.0075 & 6.4822 & 7.0807 & 0.8921 & 0.8921 & 3.0176 & 3.0181 & 0.0531 & 0.0459 & $\approx$-1 & $\approx$-1 \\
     & 0.006 & 5.7589 & 8.9159 & 0.9008 & 0.9032 & 3.0555 & 3.1066 & 0.0651 & 0.0318 & $\approx$-1 & $\approx$-1 \\
0.5  & 0.005 & 5.5643 & 10.1175 & 0.9063 & 0.9116 & 3.0758 & 3.1960 & 0.0692 & 0.0262 & $\approx$-1 & $\approx$-1 \\
     & 0.003 & 5.3091 & 13.7399 & 0.9168 & 0.9317 & 3.1106 & 3.5025 & 0.0754 & 0.0164 & $\approx$-1 & $\approx$-1 \\
     & 0.000 & 5.0633 & $\infty$ & 0.9319 & 1 & 3.1533 & $\infty$ & 0.0824 & 0 & $\approx$-1 & $\approx$-1 \\ \hline
     & 0.006 & 7.2438 & 7.8850 & 0.9045 & 0.9045 & 3.2071 & 3.2075 & 0.0463 & 0.0402 & $\approx$-1 & $\approx$-1 \\
     & 0.004 & 6.2073 & 11.3079 & 0.9168 & 0.9215 & 3.2681 & 3.3917 & 0.0609 & 0.0225 & $\approx$-1 & $\approx$-1 \\
0.2  & 0.003 & 6.0161 & 13.5101 & 0.9226 & 0.9320 & 3.2920 & 3.5600 & 0.0646 & 0.0170 & $\approx$-1 & $\approx$-1 \\
     & 0.002 & 5.8713 & 17.0251 & 0.9283 & 0.9445 & 3.3137 & 3.8414 & 0.0676 & 0.0119 & $\approx$-1 & $\approx$-1 \\
     & 0.000 & 5.6562 & $\infty$ & 0.9392 & 1 & 3.3521 & $\infty$ & 0.0727 & 0 & $\approx$-1 & $\approx$-1 \\ \hline
     & 0.005 & 7.2378 & 9.1628 & 0.9127 & 0.9130 & 3.3294 & 3.3382 & 0.0479 & 0.0320 & -1 & -1 \\
     & 0.004 & 6.7213 & 11.0569 & 0.9191 & 0.9219 & 3.3642 & 3.4384 & 0.0547 & 0.0236 & -1 & -1 \\
0.0  & 0.003 & 6.4513 & 13.3434 & 0.9253 & 0.9322 & 3.3933 & 3.5987 & 0.0591 & 0.0176 & -1 & -1 \\
     & 0.002 & 6.2633 & 16.9089 & 0.9313 & 0.9446 & 3.4191 & 3.8713 & 0.0625 & 0.0121 & -1 & -1 \\
     & 0.000 & 6.0000 & $\infty$ & 0.9428 & 1 & 3.4641 & $\infty$ & 0.0680 & 0 & -1 & -1 \\ \hline
     & 0.0045 & 7.6191 & 9.6428 & 0.9175 & 0.9178 & 3.4282 & 3.4368 & 0.0450 & 0.0300 & $\approx$-1 & $\approx$-1 \\
     & 0.004 & 7.2560 & 10.7542 & 0.9209 & 0.9223 & 3.4486 & 3.4871 & 0.0492 & 0.0251 & $\approx$-1 & $\approx$-1 \\
-0.2 & 0.003 & 6.8684 & 13.1635 & 0.9274 & 0.9324 & 3.4833 & 3.6377 & 0.0545 & 0.0181 & $\approx$-1 & $\approx$-1 \\
     & 0.002 & 6.6276 & 16.7887 & 0.9337 & 0.9448 & 3.5132 & 3.9012 & 0.0583 & 0.0124 & $\approx$-1 & $\approx$-1 \\
     & 0.000 & 6.3114 & $\infty$ & 0.9457 & 1 & 3.5645 & $\infty$ & 0.0643 & 0 & $\approx$-1 & $\approx$-1 \\ \hline
     & 0.004 & 8.2385 & 10.0723 & 0.9228 & 0.9229 & 3.5575 & 3.5629 & 0.0406 & 0.0286 & $\approx$-1 & $\approx$-1 \\
     & 0.003 & 7.4787 & 12.8613 & 0.9298 & 0.9328 & 3.6022 & 3.6969 & 0.0486 & 0.0192 & $\approx$-1 & $\approx$-1 \\
-0.5 & 0.002 & 7.1342 & 16.6001 & 0.9365 & 0.9449 & 3.6384 & 3.9463 & 0.0532 & 0.0127 & $\approx$-1 & $\approx$-1 \\
     & 0.001 & 6.9035 & 24.4625 & 0.9430 & 0.9610 & 3.6700 & 4.5016 & 0.0568 & 0.0070 & $\approx$-1 & $\approx$-1 \\
     & 0.000 & 6.7294 & $\infty$ & 0.9492 & 1 & 3.6985 & $\infty$ & 0.0599 & 0 & $\approx$-1 & $\approx$-1 \\ \hline
     & 0.003 & 8.5392 & 12.2145 & 0.9326 & 0.9334 & 3.7688 & 3.7980 & 0.0403 & 0.0214 & $\approx$-1 & $\approx$-1 \\
-1.0 & 0.002 & 7.9038 & 16.2601 & 0.9399 & 0.9452 & 3.8164 & 4.0221 & 0.0468 & 0.0134 & $\approx$-1 & $\approx$-1 \\
     & 0.001 & 7.5612 & 24.2593 & 0.9469 & 0.9611 & 3.8556 & 4.5516 & 0.0511 & 0.0071 & $\approx$-1 & $\approx$-1 \\
     & 0.000 & 7.3236 & $\infty$ & 0.9536 & 1 & 3.8899 & $\infty$ & 0.0546 & 0 & $\approx$-1 & $\approx$-1 \\
\end{tabular}
\end{ruledtabular}
\end{table*}

Moreover, similarly to Tab.~\ref{tab-isco-SdS} for the
Schwarzschild-de Sitter spacetime, some values of the
characteristic parameters of the spinning particle moving along
the ISCO and OSCO of the Schwarzschild black hole immersed in
quintessence are given in Tab.~\ref{tab-isco-quint}. Again, the
qualitative behaviour of the  characteristic parameters is similar
to the Schwarzschild-de Sitter case.

\section{Conclusion}\label{sec-conclusion}

In this paper we studied the motion of spinning particles in the
equatorial plane of non asymptotically flat, static and
spherically symmetric spacetimes. The condition of non asymptotic
flatness has been imposed to account for an accelerating
cosmological boundary to the black hole spacetime. It is well
known that such a condition at spatial infinity bears important
consequences for the orbits of test particles in the spacetime,
since, as a result of the repulsive cosmological effects,
particles have only a finite range of distances where stable
circular orbits can exist. In \cite{visser} it was shown that for
a non spinning particle around a supermassive black hole such a
range is comparable with the diameter of a galaxy. Therefore one
can not exclude that non asymptotic flatness may have some
relevance in astrophysics. Here we showed that the presence of
spin for the test particles may alter this structure and modify
the extent of the region where stable circular orbits exist.

\section*{Acknowledgments}

The work was developed under the Nazarbayev University Faculty
Development Competitive Research Grant No.~090118FD5348. The
authors acknowledge the support of the Ministry of Education of
Kazakhstan's target program IRN:~BR05236454 and Uzbekistan
Ministry for Innovative Development Grants No.~VA-FA-F-2-008 and
No.~MRB-AN-2019-29.

\label{lastpage}

\bibliography{spinning_references}

\end{document}